\documentclass[aps,prc,preprint,groupedaddress,showpacs]{revtex4}
\usepackage{graphicx}
\usepackage{bm}
\usepackage{latexsym}
\begin{document}
\title{ Effective three-body interactions in the $\alpha$-cluster model for 
the $^{12}\mathrm{C}$ nucleus }

\author{S.~I.~Fedotov}
\author{O.~I.~Kartavtsev}
\author{A.~V.~Malykh}
\affiliation{ Joint Institute for Nuclear Research, Dubna, 141980, Russia }
%\affiliation{ Bogoliubov Laboratory of Theoretical Physics, Joint
%Institute for Nuclear Research, Dubna, 141980, Russia }
%\affiliation{ Dzhelepov Laboratory of Nuclear Problems, Joint
%Institute for Nuclear Research, 141980, Dubna, Russia }

%\date{\today}

\begin{abstract}

Properties of the lowest $0^{+}$ states of $^{12}\mathrm{C}$ are calculated 
to study the role of three-body interactions in the $\alpha$-cluster model. 
An additional short-range part of the local three-body potential is introduced 
to incorporate the effects beyond the $\alpha$-cluster model. 
There is enough freedom in this potential to reproduce the experimental 
values of the ground-state and excited-state energies and the ground-state 
root-mean-square radius. 
The calculations reveal two principal choices of the two-body and three-body 
potentials. 
Firstly, one can adjust the potentials to obtain the width of the excited 
$0_2^+$ state and the monopole $0_2^+ \to 0_1^+ $ transition matrix element 
in good agreement with the experimental data. 
In this case, the three-body potential has strong short-range attraction 
supporting a narrow resonance above the $0_2^+$ state, the excited-state wave 
function contains a significant short-range component, and the excited-state 
root-mean-square radius is comparable to that of the ground state. 
Next, rejecting the solutions with an additional narrow resonance, one finds 
that the excited-state width and the monopole transition matrix element 
are insensitive to the choice of the potentials and both values exceed 
the experimental ones. 

\end{abstract}

\maketitle

\section{Introduction}

As the $\alpha$-particle is the most tightly bound nucleus, 
a variety of the low-energy nuclear properties can be successfully described 
within the framework of the $\alpha$-cluster model. 
The effective two-body and, for more than two $\alpha$-particles, at least 
three-body potentials must be determined as an input for the model. 
The three-body calculations allow one to reduce ambiguity in the two-body 
potential which could not be determined merely from the two-body data. 
In this respect, the basic problem is to check the model for the system of 
three $\alpha$-particles, thus, the effective potentials should be chosen 
by fitting the main characteristics of the $^{12}\mathrm{C}$ nucleus 
to the experimental values. 

In spite of significant simplifications provided by the $\alpha$-cluster 
model, there are complicated problems inherent in the processes with few 
charged particles in the initial or final state. 
For the problems of this kind the main difficulty stems from the necessity 
to describe the continuum wave function and even qualitative understanding 
of the reaction mechanism is crucial. 
The formation of the $^{12}\mathrm{C}$ nucleus in the triple-$\alpha$ 
low-energy collisions, which plays a key role in stellar nucleosynthesis 
\cite{Salpeter52,Hoyle54}, is a well-known example. 
More examples are double-proton radioactivity, which has been 
a subject of thorough experimental and theoretical investigations during 
the last years (for details see the recent 
reviews~\cite{Grigorenko01,Grigorenko03}) and decay of the long-lived 
$^{12}\mathrm{C} (1^{+})$ state~\cite{Fynbo03}. 
Note also a description of multi-cluster decay of atomic nuclei by using  
the quasi-classical approach to Coulomb-correlated penetration through 
a multidimensional potential barrier~\cite{Kartavtsev04}. 

In the triple-$\alpha$ reaction both the low-energy $\alpha $-$\alpha$ 
resonance (the ground state of $^8\mathrm{Be}$) and the near-threshold 
three-body resonance ($^{12}\mathrm{C} (0_2^+)$ state) play an important role. 
These resonances are predicted in Ref.~\cite{Hoyle54} as unique possibility 
for helium burning that provides the only explanation for observable abundance 
of elements in the universe. 
Due to these resonances, the triple-$\alpha$ reaction in stars goes through 
the sequential reaction $3\alpha \to {^8\mathrm{Be}} + \alpha \to 
{^{12}\mathrm{C}} (0_2^+) \to {^{12}\mathrm{C}} + \gamma$. 
The predicted $^{12}\mathrm{C} (0_2^+)$ state, starting with 
the observation~\cite{Dunbar53,Cooc57}, was thoroughly studied later on, 
in particular, the decay mechanism was investigated in Ref.~\cite{Freer94}. 
The corresponding theoretical problem is the microscopic calculation 
of the resonance width extremely small on the nuclear scale and 
the $0_2^+ \to 0_1^+ $ monopole transition matrix element (MTME). 

Among other interesting problems connected with description 
of $\alpha$-cluster nuclei, one should mention the nonresonance reaction 
$3\alpha\to{^{12}\mathrm{C}}$, which is responsible for helium burning 
at ultra-low temperatures and high densities as in accretion on white 
dwarfs and neutron stars~\cite{Cameron59}. 
Whereas a number of model calculations of the nonresonance reaction are 
available~\cite{Nomoto85,Langanke86,Fushiki87,Schramm92}, a consistent 
three-body description is needed to avoid a possible error of a few orders 
of magnitude in the calculated reaction rate. 
Note also that recently the $\alpha$-cluster states in nuclei have attracted 
attention in connection with the problem of $\alpha$-particle condensation 
(see, e. g., Ref.~\cite{Funaki05} and references therein). 

A focus of the present paper is to shed light, using the technique of 
Ref.~\cite{Fedotov04}, on the role of the three-body interactions 
in description of the lowest $0^{+}$ states of $^{12}\mathrm{C}$. 
The main question to be answered is to what extent the $\alpha$-cluster model 
is able to reproduce the experimental energies and sizes of the nuclear 
states. 
The next one, more challenging problem, is to describe the fine 
characteristics, such as the width of the near-threshold $0^+_2$ state and 
$0^+_2 \to 0^+_1$ MTME, which are sensitive to the choice of the potentials. 
In realistic calculations, the finite size of the $\alpha$-particle implies 
crucial importance of the effective three-body interactions for reliable 
description within the framework of the $\alpha$-cluster model 
\cite{Fedorov96,Filikhin00,Fedotov04}. 
Furthermore, the effective three-body interactions could be used to take into 
account the non-$\alpha$-cluster structure of the nucleus at short distances 
in addition to the effect of $\alpha$-particle distortions at large distances.
Clearly, the choice of the effective two-body and three-body potentials must 
be governed by the results of the three-body calculations aimed at optimal 
description of the $^{12}\mathrm{C}$ characteristics. 

\section{Theoretical background}

The present paper is aimed to chose, by means of microscopic three-body 
calculation of the ground and first excited $0^+$ states of $^{12}\mathrm{C}$, 
the effective three-body and two-body potentials of the $\alpha$-cluster 
model. 
It is assumed that all the effects connected with both the internal structure 
of $\alpha$-particles and the identity of nucleons are incorporated 
in the effective potentials. 
The two-body input is defined by the local $\alpha $-$\alpha$ potential that 
reproduces the experimental energy and width of the near-threshold 
$\alpha $-$\alpha$ resonance (ground state of $^8\mathrm{Be}$). 
More precisely, with the $^8\mathrm{Be}$ energy fixed and its width varying 
within the experimental uncertainty, a set of two-body potentials is 
constructed by modification of the Ali-Bodmer s-wave potential~\cite{Ali66}. 
One uses a simple, and suitable for calculation, functional form of 
the three-body potential, which depends only on the collective variable, 
viz, the hyper-radius. 
A sum of two Gaussian terms is used, which makes it possible to take into 
account both the effect of $\alpha$-particle distortions at large distances 
and the short-range non-$\alpha$-cluster effects. 
Calculation of the resonance width and the MTME makes sense only if, 
not only the ground-state energy but also the resonance position and 
the root-mean-square (rms) radius of the ground state are fixed 
at the experimental values. 
These requirements are satisfied by adjusting the parameters of the three-body 
potential.  

The technical details and the numerical procedure are basically the same as 
in the previous paper~\cite{Fedotov04}, therefore, only a sketch of the 
calculational method will be given below. 
The method is based on the expansion of the total wave function in terms 
of the eigenfunctions on a hypersphere~\cite{Macek68}. The eigenvalue problem 
on a hypersphere is numerically solved by using the variational 
method. 

The units $\hbar = m = e = 1$ are used throughout the paper unless other 
is specified. 
The scaled Jacobi coordinates are defined as 
${\mathbf x}_i = {\mathbf r}_j-{\mathbf r}_k ,\ 
{\mathbf y}_i = (2{\mathbf r}_i-{{\mathbf r}_j - {\mathbf r}_k})/\sqrt{3}$, 
where ${\bf r}_i$ is the position vector of the $i$th particle. 
The hyper-spherical variables $\rho $, $\alpha_i$, and $\theta_i $, 
are defined via the Jacobi coordinates by the relations 
$x_i = \rho \cos\frac{\alpha_i}{2}$, $y_i = \rho \sin\frac{\alpha_i}{2}$, and 
$\cos\theta_i = \frac{({\mathbf x}_i{\mathbf y}_i)}{x_i y_i} $.

The Schr\"odinger equation for three $\alpha$-particles is
\begin{equation}
\left(-\Delta_{\mathbf x} - \Delta_{\mathbf y} + \sum_{j=1}^3 V(x_j) 
+ V_3(\rho) - E \right)\Psi = 0 \ , 
\label{eq}
\end{equation} 
where the total interaction contains the pair-wise potentials $V(x_i)$ 
and the three-body potential $ V_3(\rho) $. 
The two-body potential is a sum $V(x) = V_s(x) + V_c(x) $, where 
\begin{equation}
V_s(x) = V_r e^{-\mu_{r}^{2} x^2} - V_a e^{-\mu_{a}^{2} x^2}  
\label{ef2}
\end{equation} 
and $V_c(x) = \displaystyle\frac{4}{x} $. 
The three-body potential is taken as an obvious extension of the potential 
used in Refs.~\cite{Fedorov96,Filikhin00,Fedotov04}, 
\begin{equation}
V_3(\rho) = V_0 e^{-(\rho/b_0)^2} + V_1 e^{-(\rho/b_1)^2} \ . 
\label{ef3}
\end{equation} 
With the expansion of the total wave function 
\begin{equation}
\Psi = \rho^{-5/2}\sum_n f_n(\rho) \Phi_n(\alpha,\theta,\rho)  
\label{coup}
\end{equation} 
in a series of the normalized eigenfunctions $\Phi_n$ satisfying the equation 
on the  
\begin{eqnarray}
\label{eqa} 
&& \left[ \frac{\partial^2}{\partial\alpha^2} + 
2 \cot\alpha \frac{\partial}{\partial\alpha} 
+ \frac{1}{\sin^{2}\alpha } \left(\frac{\partial^2}{\partial\theta^2} + 
\cot\theta \frac{\partial}{\partial\theta} \right) \right. - 
\nonumber \\ && \\ && \nonumber 
\left. \frac{\rho^2}{4}\sum_{j=1}^3 V\left(\rho \cos\frac{\alpha_j}{2}\right) 
+ \lambda_{n}(\rho)\right] \Phi_n(\alpha, \theta, \rho) = 0 \ , 
\end{eqnarray} 
the Schr\"odinger 
equation~(\ref{eq}) is routinely transformed to the system of hyper-radial 
equations (HRE)
\begin{eqnarray}
\label{eq3}
&&\left[ \frac{\partial^2}{\partial \rho^2} - \frac{1}{\rho^2}
\left( 4\lambda_n(\rho) + \frac{15}{4} \right) - V_3(\rho) + 
E\right] f_{n}(\rho)+ \nonumber \\ && \\ \nonumber 
&&\sum_m \left( Q_{nm}(\rho)\frac{\partial}{\partial\rho} +
\frac{\partial}{\partial\rho}Q_{nm}(\rho) - 
P_{nm}(\rho) \right) f_m(\rho) = 0 \ , 
\end{eqnarray}
\begin{eqnarray}
\label{qdef}
Q_{nm}(\rho) = \left\langle\Phi_n \biggm| 
\frac{\partial\Phi_m}{\partial\rho}\right\rangle, 
P_{nm}(\rho) = \left\langle\frac{\partial\Phi_n}{\partial\rho} \biggm| 
\frac{\partial\Phi_m}{\partial\rho}\right\rangle, 
\end{eqnarray}
where $\langle\cdot|\cdot\rangle$ stands for integration on the hyper-shere. 

The functions $\lambda_n(\rho)$, $Q_{nm}(\rho)$, and $P_{nm}(\rho)$ 
are calculated by using the variational solutions of the eigenvalue 
problem~(\ref{eqa}). 
In view of the symmetry of $\Phi_n(\alpha,\theta,\rho)$, which follows from 
the identity of $\alpha$-particles, the variational trial functions 
are chosen to be symmetric under any permutation of particles. 
Few types of trial functions are used, which provides  flexibility of 
the variational basis needed to describe an essentially different structure 
of the wave function at different values of $\rho$, in particular, the two- 
and three-cluster configurations in the asymptotic region. 
Thus, the variational basis contains a set of the symmetric hyper-spherical 
harmonics which are eigenfunctions of the differential operator 
in Eq.~(\ref{eqa}). 
Furthermore, to describe the two-cluster configuration, symmetrized 
combinations of the $\rho$-dependent two-body functions $\phi_i(x)$ 
are included in the basis. 
As in Ref.~\cite{Fedotov04}, a set of $\phi_i(x)$ includes Gaussian functions 
$\phi_i(x) = \exp{(-\beta_i x^2)} $, which allows the two-cluster wave 
function to be described within the range of the nuclear potential $V_s(r)$, 
and the function $\phi(x) = x^{1/4}\exp{(-4\sqrt{x}(1 + ax))} $ to describe 
the two-cluster wave function in the sub-barrier region. 

Solutions of the eigenvalue problem (at $E<0$) and the 
$\alpha + {^8\mathrm{Be}}$ scattering problem (at $E>0$) 
for HRE~(\ref{eq3}) provide the properties of the ground $0_1^+$ state 
and the excited $0_2^+$ resonance state, respectively.  
The resonance position $E_r$ and width $\Gamma$ are determined by fitting 
the scattering phase shift $\delta_E$ to the Wigner dependence on energy
\begin{equation} 
\label{res1}
\cot(\delta_E - \delta_{bg}) = \frac{2}{\Gamma}(E_r - E) \ , 
\end{equation}
where the background phase shift $\delta_{bg}$ is of no interest 
for the present calculation. It is suitable to treat the ultra-narrow 
$0_2^+$ resonance state on equal footing with the ground state.  
Therefore, its wave function, defined as the scattering solution 
at the resonance energy $E_r$, is normalized on the finite interval 
$0 \le \rho \le \rho_t $, where $\rho_t $ is the turning point 
of the first-channel hyper-radius potential 
$U_1(\rho) = \frac{1}{\rho^2}\left( 4\lambda_1(\rho) + \frac{15}{4} 
\right) + V_3(\rho) + P_{11}(\rho)$. 
Thus, the rms radii $R^{(i)}$ of the ground ($i = 1$) and excited ($i = 2$) 
states and MTME $M_{12}$ are defined by the expressions 
\begin{equation}
\label{rms2}
R^{(i)} = \sqrt{R^2_{\alpha} + \frac{1}{6} \bar\rho^2_i } \ , \qquad
\bar\rho^2_i = \sum_{n}\int\limits_0^{\infty}\left|
f^{(i)}_n(\rho)\right|^2\rho^2 d\rho \ , 
\end{equation}
where $R_{\alpha} = 1.47$ fm is the rms radius of the $\alpha$-particle, and 
\begin{equation}
\label{tranme2}
M_{12} = \sum_{n}\int\limits_0^{\rho_t}f^{(2)}_n(\rho)f^{(1)}_n(\rho)\rho^2 
d\rho \ . 
\end{equation}

\section{Results}

Calculations have been performed with a family of the two-body 
$\alpha $-$\alpha$ potentials $V_s$~(\ref{ef2}), which are obtained 
by modification of potential (a) from Ref.~\cite{Ali66}. 
With the ranges of the repulsive and attractive parts fixed at the values 
$\mu_r^{-1} = 1.53 fm$ and $\mu_a^{-1} = 2.85 fm$, the parameters $V_r$ and 
$V_a$ were chosen to reproduce the experimental energy 
$E_{2\alpha} = 91.89$ keV~\cite{Ajzenberg88} of the $\alpha $-$\alpha$ 
resonance (ground state of $^8\mathrm{Be}$) and to vary its width within 
the experimental uncertainty $\gamma = 6.8 \pm 1.7$ eV~\cite{Ajzenberg88}. 
As the width $\gamma $ unambiguously determines the parameters of the two-body 
potential, in the following the potential will be marked by $\gamma $. 
A partial set of the parameters $V_r$ and $V_a$ and the widths $\gamma$ 
is presented in Table~\ref{tab1}. 
\begin{table}[thb]
\caption{Parameters of the $\alpha $-$\alpha$ potential 
$V_s$~(\protect\ref{ef2}) providing the $\alpha $-$\alpha$ resonance 
widths $\gamma$. }
\label{tab1}
\begin{tabular}{ccc}
%\hline \hline 
$\gamma $(eV) & $V_r$(MeV) & $V_a$(MeV)  \\
\hline
  5.69 & 35.024    & 19.492 \\
  6.20 & 52.772    & 22.344 \\ 
  6.37 & 60.051    & 23.359 \\
  6.40 & 61.220	   & 23.516 \\
  6.50 & 66.028	   & 24.141 \\ 
  6.60 & 71.057	   & 24.766 \\   
  6.80 & 82.563    & 26.1 
\end{tabular}
\end{table} 

The three-channel system of HREs~(\ref{eq3}) is solved to calculate 
the ground- and excited-state energies $E_{gs}$ and $E_r$, the rms radii 
$R^{(i)}$, the excited-state width $\Gamma$, and the monopole transition 
matrix element $M_{12}$. 
Convergence in a number of HRE is sufficiently fast and solution of three HRE 
allows the resonance width to be determined with an accuracy not worse than 
$1$ eV. 
Generally, the parameters of the numerical procedure and an accuracy 
of the calculated $E_{gs}$, $E_r$, $\Gamma$, $R^{(i)}$, and $M_{12}$ were 
the same as in Ref.~\cite{Fedotov04}. 
Using the numerical procedure of determination of $E_{gs}$, $E_r$, 
and $R^{(1)}$, the parameters of the three-body potential for each two-body 
potential were determined by solving the nonlinear inverse problem of fixing 
the ground- and excited-state energies and the excited-state rms radius at 
the experimental values $E_{gs} = -7.2747$ MeV, $E_r = 0.3795$ MeV 
\cite{Ajzenberg90}, and $R^{(1)}_{exp} = 2.48\pm 0.22$ fm 
\cite{Ruckstuhl84,Offermann91}. 

At the first stage of the calculations only the one-term potential ($V_1 = 0$) 
was studied for better understanding of the dependence on the three-body 
potential $V_3(\rho)$~(\ref{ef3}). 
For this two-parameter potential, only $E_{gs}$ and $E_r$ are fixed at 
the experimental values to determine $V_0$ and $b_0$. 
The calculation gives two types of solutions, that is, two families of 
one-term three-body potentials, whose parameters $V_0$ and $b_0$ are 
presented in Table~\ref{tab2}. 
For one type of solutions, three-body potentials are rather extended 
with the range about $b_0 = 4.5$ fm and strength $|V_0| < 40$ MeV. 
The ground-state rms radius is in the range $2.2 \ \mathrm{fm} < R^{(1)} < 2.8 
\ \mathrm{fm} $, which includes the experimental value, 
whereas $\Gamma$ and $M_{12}$ 
significantly exceed the experimental values $\Gamma = 8.5 \pm 1.0$ eV and 
$M_{12} = 5.48 \pm 0.22$ fm$^2$~\cite{Ajzenberg90}.  
For another type of solutions, $b_0$ is about twice as small and $|V_0| $ 
exceeds $80$ MeV. 
The ground-state rms radius is lower than $R^{(1)}_{exp}$, nevertheless, 
$\Gamma$ and $M_{12}$ are in better agreement with experiment than 
in the previous case. 
As the ground-state size $R^{(1)}$ cannot be fixed at the experimental value 
by using the one-term three-body potential, it is not surprising that finer 
properties $\Gamma$ and $M_{12}$ vary in a wide range with variations 
of the two-body potential. 
\begin{table}[hbt]
\caption{Two families of solutions with the one-term three-body potential 
($V_1 = 0$) for a number of two-body potentials marked by the widths 
$\gamma$ (eV) of the $\alpha $-$\alpha$ resonance. 
Shown are the parameters $b_0$ (fm) and $V_0$ (MeV), rms radii $R^{(i)}$ (fm), 
width of the excited state $\Gamma$ (eV), and monopole transition matrix 
element $M_{12}$ (fm$^2$). }
\label{tab2}
\begin{tabular}{ccccccccccccccc}
$\gamma$ & \phantom{w} & $b_0$& $V_0$ & $\Gamma$ & $R^{(1)}$ & $R^{(2)}$ 
& $M_{12}$ 
& \phantom{w} & $b_0$& $V_0$ & $\Gamma$ & $R^{(1)}$ & $R^{(2)}$ & $M_{12}$ \\ 
%\hline 
\cline{1-1} \cline{3-8} \cline{10-15} 
 5.69 &     & 4.5001 & -18.600 & 13.0 & 2.35 & 3.7 & 8.59 
  & & 2.2310 & -89.941 & 8.2 & 2.02 & 3.4 & 6.46 \\
 6.20 &     & 4.6006 & -20.824 & 15.9 & 2.45 & 3.8 & 8.87
  & & 2.3314 & -113.28 &  9.7 & 2.09 & 3.5 & 6.90  \\
 6.37 &     & 4.6247 & -21.643 & 16.9 & 2.48 & 3.9 & 8.93
 & & 2.3472 & -125.05 & 10.2 & 2.12 & 3.5 & 7.01  \\
 6.40 &     & 4.6455 & -21.640 & 17.2 & 2.48 & 3.9 & 8.97
  & & 2.3464 & -127.48 & 10.4 & 2.12 & 3.5 & 7.03  \\
 6.50 &     & 4.6379 & -22.297 & 17.6 & 2.50 & 3.9 & 8.97 
  & & 2.3547 & -135.38 & 10.7 & 2.13 & 3.5 & 7.09 \\
 6.60 &     & 4.6455 & -22.838 & 18.1 & 2.51 & 3.9 & 8.99 
 & & 2.3584 & -144.47 & 11.0 & 2.14 & 3.6 & 7.13 \\
 6.80 &     & 4.6531 & -24.047 & 19.3 & 2.55 & 4.0 & 9.03 
  & & 2.3611 & -166.39 & 11.7 & 2.17 & 3.6 & 7.22 
\end{tabular}  
\end{table}     

The results for the one-term potential ($V_1 = 0$) clearly show lack 
of simultaneous description for the ground-state size $R^{(1)}$ and 
the excited-state characteristics $\Gamma$ and $M_{12}$. 
As far as it does not seem reasonable to improve agreement between 
calculation and experiment for the very fine properties $\Gamma$ and $M_{12}$ 
at the expense of the ground-state rms radius, one concludes that the one-term 
three-body potential is too simple to describe the real nucleus. 
One can readily propose to contaminate both the short-range and the long-range 
term in the three-body potential to obtain compromising description 
of the ground-state and excited-state characteristics. 

\begin{figure}[htb]
\caption{Calculated $M_{12}$--$\Gamma$ relations. Each line depicts the 
result for the two-body potential marked by the two-body resonance width 
$\gamma$. 
The point with errorbars shows the experimental data. 
The corresponding $R^{(2)}$--$\Gamma$ relations are shown in the inset.}
\includegraphics[angle=0, width=0.9\textwidth]{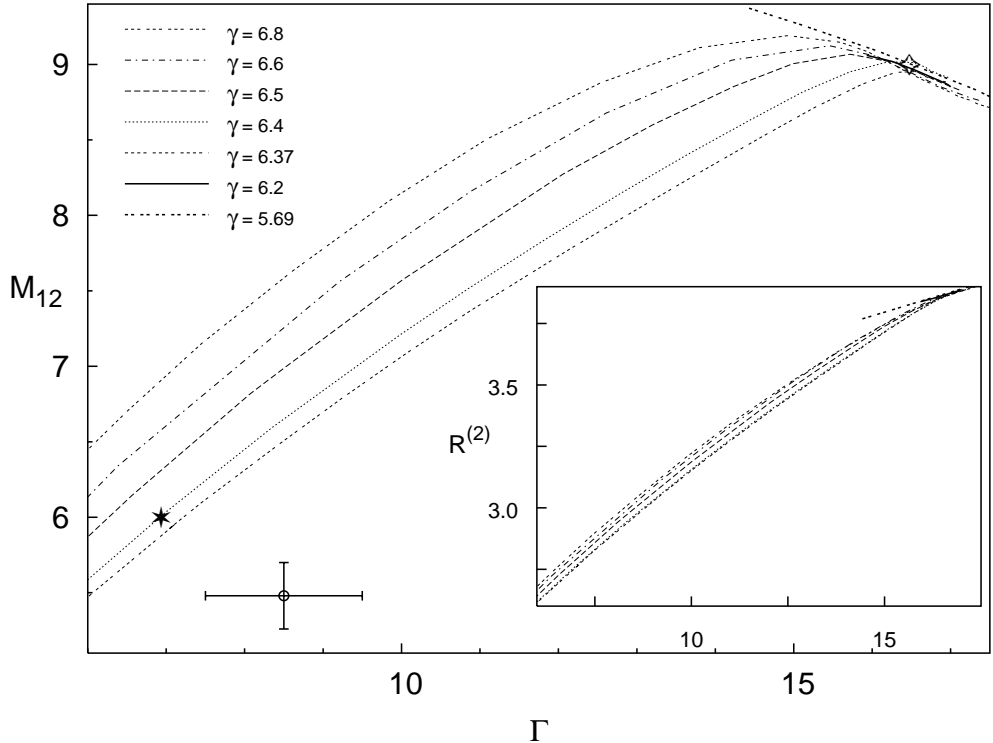}
\label{fig1}
\end{figure}

At the main route of calculations, four parameters of the three-body 
potential $V_3(\rho)$~(\ref{ef3}) are used to fix the basic properties, viz., 
the ground-state and excited-state energies and the ground-state rms 
radius at the experimental values~\footnote{In the present calculation, 
the ground-state rms radius $R^{(1)}$ is fixed at the elder experimental value 
$2.47$ fm~\protect\cite{Reuter82} that does not reflect on the conclusions. }. 
Varying one remaining degree of freedom in the four-dimensional space 
of parameters $V_{0, 1}, b_{0, 1} $ of the three-body potential, one obtains 
a one-parameter set of solutions, which is suitably represented for each 
two-body potential by a line in the $\Gamma $--$M_{12}$ plane, as shown 
in Fig.~\ref{fig1}. 
It turns out that some of the calculated potentials, namely, those for which 
the parameter $b_0 > 6$ fm, are of the form of a shallow well with a long 
tail. 
These solutions of unreasonably long range were withdrawn from 
the consideration and will not be presented. 

\begin{figure}[thb]
\caption{The first-channel hyper-radial potentials $U_1(\rho)$ calculated 
for the two-body potential providing $\gamma = 6.4$ eV. 
Dash-dotted and dotted lines depict $U_1(\rho)$ for the three-body potentials 
whose parameters are marked in Table~\protect\ref{tab3} by a diamond and 
an asterisk, respectively. 
Full and dashed lines depict $U_1(\rho)$ for the one-term three-body potential 
whose parameters are given in the line 4 of Table~\protect\ref{tab2}. 
The inset shows the first-channel radial functions $f_1(\rho )$ 
for the potential marked by an asterisk (full and dash-dotted lines for 
the ground and excited states) and for the potential marked by a diamond 
(dashed and dotted lines for the ground and excited states). }
\includegraphics[angle=0, width=0.9\textwidth]{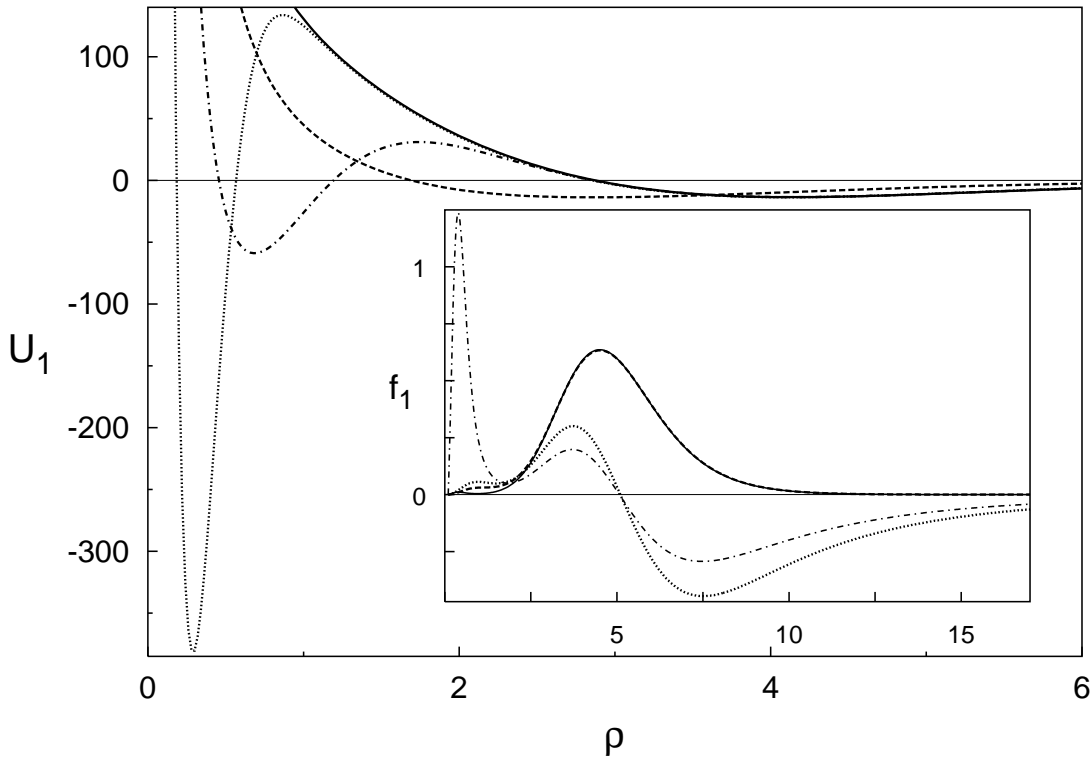}
\label{fig2}
\end{figure}

Furthermore, the parameters of the long-range term in the three-body 
potential $V_0 \approx 20$ MeV and $b_0 \approx 4.5$ fm are similar to those 
found in the calculations with the one-term potential. 
Thus, the long-range tails of the four-parameter three-body potentials and one 
of the one-term potentials practically coincide. 
On the other hand, the one-term potentials of another type look like an 
average of the full three-body potentials at short distances. 
These qualitative features are seen in Fig.~\ref{fig2}, where 
the first-channel hyper-radial potentials $U_1(\rho) $ are presented. 

All the solutions turn out to pass through a small common area about 
$\Gamma \approx 16.5$ eV and $M_{12} \approx 9$ fm$^2$, which is marked 
by a diamond in Fig.~\ref{fig1}. 
This area is well separated from the experimental values. 
Correspondingly, the calculated values of the excited-state rms radius are 
concentrated around the value $R^{(2)} \approx 3.9$ fm.  
This surprising insensitivity of $\Gamma $, $M_{12}$, and $R^{(2)}$ to 
the choice of the two-body and three-body potentials results from the imposed 
requirement to fix the ground-state rms radius at the experimental value. 

The solutions could be separated in two classes, which are characterized 
by a sign of the short-range term in the three-body potential. 
The solutions of the first class ($V_1 < 0$) are found for $\gamma > 6.35$ eV 
and those of the second class ($V_1 > 0$) for $\gamma < 6.35$ eV. 
Note that this separation is correlated with the dependence of 
the ground-state rms radius $R^{(1)}$ on $\gamma $ found in the above 
calculations for the extended one-term potentials. 
If $R^{(1)} < R^{(1)}_{exp}$, which takes place for $\gamma < 6.35$ eV, 
one needs to add a repulsive term ($V_1 > 0$), and if 
$R^{(1)} > R^{(1)}_{exp}$ (for $\gamma > 6.35$ eV), an attractive term 
must be added to fix $R^{(1)}$ at the experimental value. 
For the second-class solutions, the larger $\Gamma $ the smaller $M_{12}$, 
therefore, the corresponding lines never approach the experimental data. 
On the contrary, for the solutions of the first class ($\gamma > 6.35$ eV), 
the lines in the $\Gamma $--$M_{12}$ plane bend downward inside the common 
area that provides an option to diminish simultaneously $\Gamma $ and 
$M_{12}$. 

More detailed consideration of the dependence on the two-body potential 
(on the parameter $\gamma $) shows that the lines in the $\Gamma $--$M_{12}$ 
plane representing the solutions of the first class form a band, as seen 
in Fig.~\ref{fig1}. 
The upper and lower borders of the band correspond to $\gamma \approx 6.8$ eV 
and $\gamma \approx 6.35$ eV, respectively. 
The dependence on $\gamma $ (for $\gamma > 6.8$ eV) becomes weak so that 
the lines in the $\Gamma $--$M_{12}$ plane are rather close to the upper 
border, though being inside the band. 
For decreasing $\gamma $ below $6.8$ eV, the lines in the $\Gamma $--$M_{12}$ 
plane shift downward until the critical value about $\gamma \approx 6.35$ eV 
is reached. 
An abrupt transition to the second class solutions takes place at the critical 
value of $\gamma $, beyond which the lines in the $\Gamma $--$M_{12}$ plane 
always remain near the common area ($\Gamma \approx 16.5$ eV, 
$M_{12} \approx 9$ fm$^2$). 
The dependence of $R^{(2)}$ on $\gamma $ is illustrated in the inset 
in Fig.~\ref{fig1}, where it is seen that for $\gamma > 6.35$ eV the lines lie 
within a narrow band in the $\Gamma $ -- $R^{(2)}$ plane. 
Alternatively, for $\gamma < 6.35$ eV, the calculated values are in a small 
area about $\Gamma \approx 16.5$ eV and $R^{(2)}\approx 3.9$ fm.  

The described features are closely connected with the form of the three-body 
potential, in particular, the smaller $\Gamma $, $M_{12}$, and $R^{(2)}$ 
the larger the strength $|V_1|$ of the attractive term and the smaller its 
range $b_1$. 
To exemplify these considerations, let us consider a typical two-body 
potential with $\gamma = 6.4$ eV and a particular three-body potential, 
which gives $\Gamma $ and $M_{12}$ (marked by an asterisk in Fig.~\ref{fig1}) 
sufficiently close to the experimental data. 
The parameters of the three-body potential, $\Gamma $, $M_{12}$, and $R^{(2)}$ 
at this point are compared in Table~\ref{tab3} with the corresponding values, 
which are typical of the common area (marked by a diamond). 
\begin{table}[htb]
\label{tab3} 
\caption{Parameters of the three-body potential and characteristics of the 
$^{12}\mathrm{C} (0^{+})$ states. 
The two-body potential provides the $\alpha$-$\alpha$ resonance widths 
$\gamma = 6.4$ eV.
An asterisk and a diamond mark two solutions which are also depicted 
in Fig.~\protect{\ref{fig1}}. 
}
\begin{tabular}{cccccccc}
& $ V_0 $ (MeV) & $ b_0 $ (fm) & $ V_1 $ (MeV) & $ b_1 $ (fm) & 
$ \Gamma $ (eV) & $ M_{12}$ (fm$^2$) & $R^{(2)}$ (fm) \\
\hline 
$\Diamond$ & -22.189 & 4.5699 & -411.719 &  1.0155 & 16.5 &  9.01 & 3.86 \\
$\ast$ &-22.867 & 4.5109 &  -1710.00 & 0.41009 & 7.0 & 6.0 & 2.76 
\end{tabular}
\end{table}
The excited-state rms radius $R^{(2)}$, as shown in the inset 
in Fig.~\ref{fig1}, decreases with decreasing $\Gamma $ from the typical value 
$R^{(2)} \approx 3.9$ fm for solutions near the common area to 
$R^{(2)} \approx 2.8$ fm for solutions near the point marked by an asterisk 
that only slightly exceeds the ground-state rms radius $R^{(1)} = 2.47$ fm. 
Diminishing of $R^{(2)}$ to these small values underlines a comparatively 
compact structure of the excited state. 
Indeed, the excited-state wave functions, as shown in the inset 
in Fig.~\ref{fig2}, are quite different for the solutions marked 
by a diamond and an asterisk. 
Nevertheless, the ground-state wave functions are surprisingly similar 
to each other. 

A drastic modification of the excited-state wave function at short distances 
for the solution with small $\Gamma $ and $M_{12}$ hints that the short-range 
attractive well in the three-body potential is able to support a near-lying 
resonance state. 
Indeed, the calculations reveal an additional resonance, whose energy changes 
from about 0.5 MeV for the solutions with small $\Gamma $ and $M_{12}$ 
to about 1 MeV for the solutions providing $\Gamma $ and $M_{12}$ near 
the common area. 
Correspondingly, the resonance width increases from hundreds of eV to hundreds 
of keV. 

\section{Summary and discussion}

The lowest $0^{+}$ states of $^{12}\mathrm{C}$ are calculated to study 
the role of the three-body interactions in the $\alpha$-cluster model. 
The method used in the present paper provides an accurate calculation 
of fine characteristics of ${^{12}\mathrm{C}}$, viz., the extremely narrow 
width $\Gamma$ of the $0^{+}_2$ state and the $0^{+}_2 \to 0^{+}_1$ MTME 
$M_{12}$. 
The two-body potentials, obtained by modification of the Ali-Bodmer potential, 
provide the exact energy of the $\alpha $-$\alpha$ resonance 
(the $^{8}\mathrm{Be}$ nucleus) while its width is allowed to vary within 
the experimental uncertainty. 
A simple two-Gaussian form of the three-body potential is chosen on 
the assumption that the potential must take into account the effects beyond 
the $\alpha$-cluster model. 
The experimental values of the ground- and excited-state energies and 
the ground-state rms radius are used to impose three restrictions on four 
parameters of the three-body potential. 
The remaining degree of freedom provides the one-parameter dependence 
of the width of the near-threshold $0^+_2$ state and the $0^+_2 \to 0^+_1$ 
MTME, which are experimentally available. 
It should be emphasized that the determination of the parameters of 
$V_3(\rho)$ by fixing $E_{gs}$, $E_r$, and $R^{(1)} $ at the experimental 
values leads to rather complicated dependence of $\Gamma$, $M_{12}$, 
and $R^{(2)} $ on the $\alpha $-$\alpha$ interactions. 

The calculations reveal that for all the two-body potentials under 
consideration $\Gamma $ and $M_{12}$ take the values about 16.5 eV and 
9.0 fm$^2$ and become essentially independent of the choice of the three-body 
potential. 
At the same time, the excited-state rms radius $R^{(2)}\approx 3.9$ fm 
noticeably exceeds the ground-state rms radius $R^{(2)}=2.47$ fm.
Both $\Gamma $ and $M_{12}$ are well above the experimental data, which 
reflects a general trend for these values to be overestimated in calculations. 
Alternatively, for the two-body potentials corresponding to $\gamma > 6.35 $ 
eV, i.~e., for the  three-body potential with a strong attractive short-range 
term, both $\Gamma $ and $M_{12}$ decrease as the strength of the attractive 
term $|V_1|$ increases and its range $b_1$ decreases. 
The solutions of this kind optionally give the values of $\Gamma $ and 
$M_{12}$ which are surprisingly close to the experimental data. 
For these solutions, the excited-state structure undergoes a considerable 
modification by a strong short-range attractive potential, which entails 
on a considerable amplification of the short-range component 
of the excited-state wave function and, hence, a decrease in the rms radius 
to unexpectedly small values $R^{(2)} \approx 2.8$ fm. 
Against intuition, the short-range component of the ground-state wave 
function decreases. 
In addition, the attractive short-range term of the three-body potential
leads to appearance of a narrow resonance above the $0^+_2$ state.

In conclusion, a family of the effective potentials was found, which allows 
the experimental values for the basic characteristics of 
the ${^{12}\mathrm{C}} (0^{+})$ states, i.~e., $E_{gs}$, $E_r$, and $R^{(1)}$, 
to be reproduced within the framework of the $\alpha$-cluster model. 
Concerning the fine characteristics, such as $\Gamma$, $M_{12}$, 
and $R^{(2)}$, the calculations reveal two principal choices of the effective 
potentials. 
For the first one, the calculated $\Gamma$ and $M_{12}$ are localized in 
small areas $\Gamma \approx 16 \pm 1$ eV and $ M_{12} \approx 9 \pm 0.5$ 
fm$^2$, noticeably above the experimental data. 
In other words, if the size of the ground state is fixed, it imposes 
a stringent constraint on the finer properties, i.e., $\Gamma$ and $M_{12}$ 
for quite arbitrary potentials.
For the second one, with strong short-range attraction supporting 
an additional narrow resonance, both $\Gamma$ and $M_{12}$ take a wide range 
of values which might be chosen near the experimental data. 
These solutions exist if a narrow resonance is allowed, however, there are no 
experimental indications of a narrow resonance above the $0^+_2$ state. 
Qualitative conclusion is that if $E_{gs}$, $E_r$, and $R^{(1)}$ are fixed at 
the experimental values, a considerable short-range component of the wave 
function is needed to improve agreement with experiment for $\Gamma$ and 
$M_{12}$. 
Certainly, the problem of reliable description of $\Gamma$ and $M_{12}$ in the 
$\alpha$-cluster model deserves a thorough investigation, e.~g., by using 
the non-local three-body potential describing a coupling with twelve-nucleon 
channel at short distances. 

\bibliography{3alpha}

\end{document}